\documentclass{aa} 
\usepackage{epsfig,latexsym,graphicx,epsf,subfigure}
\usepackage{natbib}
\newcommand{\om}{\Omega_{\rm M}}
\newcommand{\ola}{\Omega_X}
\newcommand{\ok}{\Omega_k}

\newcommand{\sm}{\cal M}
\newcommand{\chisq}{\chi^2}
\newcommand{\eqref}[1]{(\ref{#1})}
\newcommand{\snap}{\textsc{snap}}
\newcommand{\nfw}{\textsc{nfw}}
\newcommand{\ml}{\textsc{ml}}

\def\lsim{\raise0.3ex\hbox{$<$}\kern-0.75em{\lower0.65ex\hbox{$\sim$}}}
\def\gsim{\raise0.3ex\hbox{$>$}\kern-0.75em{\lower0.65ex\hbox{$\sim$}}}

\begin{document}
\title{Correcting for lensing bias in the Hubble diagram}

\author{R.~Amanullah \and
        E.~M\"ortsell \and
        A.~Goobar}
\authorrunning{Amanullah et al.}

\offprints{R.~Amanullah , rahman@physto.se}

\institute{
Physics Department, Stockholm University, 
SCFAB, S-106 91 Stockholm, Sweden} 

\date{Received ...; accepted ...}

\abstract{Gravitational lensing will cause a dispersion in the Hubble diagram
for high redshift sources. This effect will introduce a bias in the
cosmological parameter determination using the distance-redshift
relation for Type Ia supernovae. In this note we show how one can
diagnose and correct for this bias when doing precision cosmology with
supernovae.
}

\maketitle

\section{Introduction}\label{sec:intro}
During the last decade, gravitational lensing has become one of the
most important tools in cosmology.  Weak lensing measurements are used
to obtain information on the amount and distribution of matter in the
universe \citep[see, e.g.,][]{weaklensing,hennawi}. Strong lensing has
long been recognized as a tool to constrain cosmological parameters
\citep[see, e.g.,][for measurements of the Hubble constant,
$H_0$]{koopmans99,browne}. Controversial upper limits on the amount of
dark energy have been derived from the statistics of multiply imaged
quasars and radio sources \citep{kocha96,falco98}. Using a different
set of Schechter parameters for E/S0 galaxies \citet{chiba} conclude
that the same quasar data set is best fitted with a universe dominated
by dark energy, i.e., in agreement with the concordance cosmology,
$(\om,\ola)=(0.3,0.7)$ whereas \citet{keeton} argue that when
calibrated to counts of distant galaxies, the method loses most of its
sensitivity to the amount of dark energy.

Further, in \citet{multi} it is shown how one could use multiple
images of supernovae (SNe) to simultaneously measure $H_0$, the
fractional energy density components and the equation of state
parameter of a possible dark energy component.  The classical SN
cosmological test however, is to study the distance-redshift relation
using Type Ia SNe as standard candles.  The effect of gravitational
lensing on these measurements is to cause an additional dispersion in
the observed magnitudes and thus be a source of systematic error in
the cosmological parameter determination \citep[see
e.g.][]{wambsganss,sn97ff}. In this note we show how the negligence of this
effect may lead to a systematic underestimation of $\om$. We also
investigate how this bias can be diagnosed and corrected for by using
the correct probability distribution function (pdf) for the dispersion
in SN magnitudes.

In Sects.~\ref{sec:method} and \ref{sec:magdist} we discuss the
method and the numerical simulations used in the analysis. In
Sect.~\ref{sec:fit}, we investigate what kind of bias we expect if not
taking lensing effects into consideration, i.e., assuming a Gaussian
magnitude distribution.  In Sect.~\ref{sec:para}, the dispersion in
magnitudes due to gravitational lensing is parametrized for different
matter distributions and redshifts and in Sect.~\ref{sec:corr} we
correct for lensing effects by performing a maximum likelihood
(\ml) analysis using the correct pdf's.

\section{The distance-redshift cosmological test}\label{sec:method}
The distances and redshifts of Type Ia SNe can be used to fit, e.g.,
the mass energy density, $\om$, and the dark energy density, $\ola$, in a
Friedmann-Lema\^{\i}tre universe. The relation between the parameters
can be expressed as
\begin{eqnarray}
    m(\vec{\theta},{\sm},z)&=&
    {\sm}+5\log_{10}\left[d'_L(\vec{\theta},z)\right]\,,\label{eq:magdl}\\
    {\sm}&=&25+M+5\log_{10}(c/H_0)\,,\label{eq:Mscript}
\end{eqnarray}
where $\sm$ is a nuisance parameter containing the absolute magnitude,
$M$, of the SNe and $H_0$ is the Hubble constant. Here $\vec{\theta}$
represents the parameter vector $\vec{\theta}=(\om,\ola,w)$. Further
the luminosity distance $d'_L$ is given by
\begin{eqnarray}\label{eq:lumdist}
    d'_L&=&\left\{
    \begin{array}{ll}
      (1+z)\frac{1}{\sqrt{-\ok}}\sin(\sqrt{-\ok}\,I)\,, &
      \ok<0\\
      (1+z)\,I\,, & \ok=0\\
      (1+z)\frac{1}{\sqrt{\ok}}\sinh(\sqrt{\ok}\,I)\,, &
      \ok>0\\
    \end{array}
    \right. \\
    \ok&=&1-\om-\ola\,,\\
    I&=&\int_0^z\,\frac{dz'}{H'(z')}\,,\\
    H'(z)&=& 
    \sqrt{(1+z)^3\,\om+
    \ola(1+z)^{3(w+1)}+(1+z)^2\,\ok}\, .\nonumber
\end{eqnarray}
The usual way to proceed is to make a $\chisq$-fit of $(\om, \ola,
\sm)$ using Eq.~\eqref{eq:magdl} assuming that the dispersion in
magnitudes is Gaussian distributed (as the intrinsic dispersion of
Type Ia SNe seems to be). The same procedure can be used to fit, e.g.,
the equation of state parameter, $w$, of a more general dark energy
component \citep{goliath}.

Gravitational lensing will however induce an asymmetry in the
magnitude distribution and in order to avoid a bias in the
cosmological parameter estimation, we need to perform a {\ml}-analysis
with the correct pdf.

\section{Simulated data sets}\label{sec:magdist}
We use the numerical simulation package, SNOC, the SuperNova
Observation Calculator \citep{code_aa}, to obtain simulated samples of
the intrinsic dispersion and gravitational lensing effects of Type Ia
SNe. The intrinsic dispersion and measurement error is represented by
a Gaussian distribution with $\sigma_m=0.16$ mag. Gravitational
lensing effects are calculated by tracing the light between the source
and the observer by sending it through a series of spherical cells in
which the dark matter distribution can be specified.

We model compact dark matter as point masses and smooth dark matter
with the Navarro-Frenk-White [\nfw; \citet{NFW}] density profile using
halo parameters, mass distributions and number densities as outlined
in \citet{MGEM-lens}. Note that the results obtained in this paper are
not sensitive to the exact parameterization of the smooth halo
profile.  Also, the results are independent of the individual masses
of the compact objects as well as their clustering properties on
galaxy scales \citep{HW98,MGEM-lens}. Any eventual small subhalo
structure in the dark matter halos does not act as a compact component
\citep{fraction}.


Two different types of simulated data sets have been used for the
analysis presented in this note. The first one, A, assumes that the
number of observed Type~Ia SNe is constant per comoving volume, which
means that the number of events increases rapidly for higher
redshifts. The B~distribution instead assumes a constant number of SNe
per redshift interval, i.e., a uniform $z$-distribution. Each data set
consists of 2000 SNe in the redshift interval $0.01<z<2.0$.

The number 2000 corresponds roughly to one year's data from the
proposed satellite telescope Supernova/Acceleration Probe [\snap;
\citet{snap}].
The exact redshift distribution of the Type Ia SNe to be followed by
\snap\ is the subject of several ongoing science and instrumental
optimization studies. The current anticipated distribution is roughly
a constant rate per comoving volume for $z<1$ and a uniform
distribution at higher redshifts. In that respect is the
A~distribution to be regarded as a limiting case of maximal lensing
effects due to the large number of high redshift SNe.

To each of these data sets, 300~SNe between $0.04<z<0.08$ have been
added. These numbers correspond to the predicted results from the
Supernova Factory Campaign~\citep{snfac} that is scheduled to start in
2003. Since lensing effects get larger at higher
redshifts, the magnitudes of theese SNe are unaffected. For example,
for a supernova at redshift $z=0.08$ the lensing effects are in
general of the order of 0.01 magnitudes for a dark matter model where
the fraction of compact objects is assumed to be 20\ts\%.

All simulations have been made assuming a universe with $\om=0.3$,
$\ola=0.7$, $w=-1$ and $H_0=65$~km\,s$^{-1}\,$Mpc$^{-1}$.

\section{The $\chisq$-fit of cosmological parameters}\label{sec:fit}
\begin{figure}[thb]
  \centering
  \subfigure[Fits in the $(\om,\ola)$-plane.]{\label{fig:omolbias}
    \hbox{\epsfig{figure=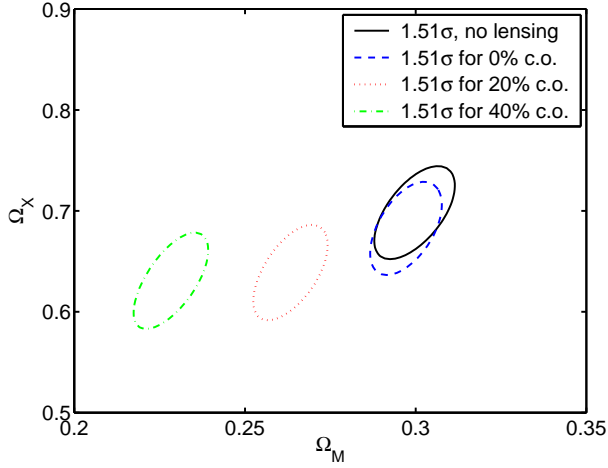,width=.45\textwidth}}}\\
  \subfigure[Fits in the $(\om,w)$-plane assuming a flat universe.]{%
    \label{fig:omwbias}%
    \hbox{\epsfig{figure=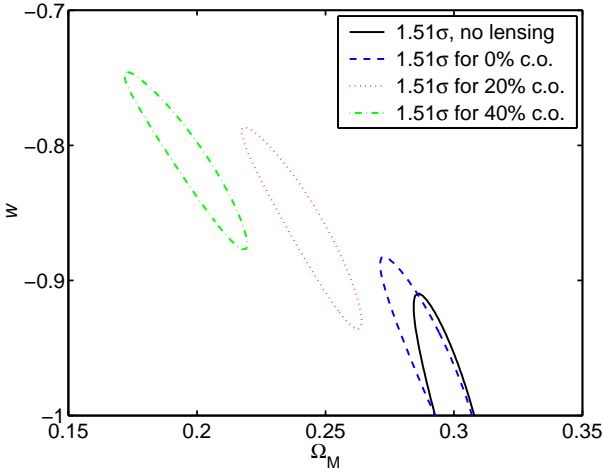,width=.45\textwidth}}}
  \caption{Confidence contours showing the 1.51\,$\sigma$ level, i.e.,
    the ~68\ts\% level for two parameters, for three-parameter
    $\chisq$-fits where the third parameter $\sm$ is treated as a
    nuisance parameter. All fits are based on a simulated sample of
    2300~SNe according to the data set~B, but different gravitational
    lensing models have been considered. The dashed contours show the
    fit when a {\nfw} halo model has been applied, and the dotted and
    dash-dotted contours represent different fractions of compact
    objects (c.o.) in the dark matter model.}
  \label{fig:bias}
\end{figure}

Making a $\chisq$-fit of $(\om, \ola, \sm)$ using Eq.~\eqref{eq:magdl}
(assuming $w=-1$) will for the data set~B, described in
Sect.~\ref{sec:magdist}, result in the solid contour presented in
Fig.~\ref{fig:omolbias} if lensing effects are absent.
If lensing dispersion is added, the fits of the cosmological
parameters will be biased in accordance with the figure. All contours
show the $1.51\,\sigma$ level, which corresponds to a $\sim68\,\%$
confidence region of including the true value of the two parameters.

The amount of bias depends on the matter distribution responsible for
the lensing effects. The dashed contour shows the result if only a {\nfw}
halo model is considered, while the dotted (dash-dotted) contour is
based on a simulation where the fraction of compact objects in the
dark matter model is 20\ts\% (40\ts\%).

Fig.~\ref{fig:omwbias} shows similar fits for the $(\om,w)$-plane
where a flat universe is assumed $(\ola=1-\om)$. $\sm$ is still
treated as a nuisance parameter, i.e., no prior knowledge is assumed.
Also this figure is based on the data set~B, but using the
A~distribution does not significantly alter the qualitative nature of
the results. Gravitational lensing starts to become important for
$z>1.0$, and in the interval $1.0<z<2.0$, the distributions A and B
are quite similiar.

In Fig.~\ref{fig:omwbias} it is evident how important it is to
consider lensing effects since a naive analysis of lensed data may in
fact misleadingly exclude a cosmological constant as the explanation
for the dark energy.

The size of the bias, i.e., the systematic errors due to lensing in
the above figures depends on the asymmetry of the magnitude
distribution, which is described in Sect.~\ref{sec:disp}. This in turn
depends on the redshift for a given dark matter model. The statistical
error -- the sizes of the ellipses -- on the other hand only depend on
the sample size so it is interesting to see where these two errors are
comparable.

\begin{figure}[thb]
  \centerline{\hbox{\epsfig{figure=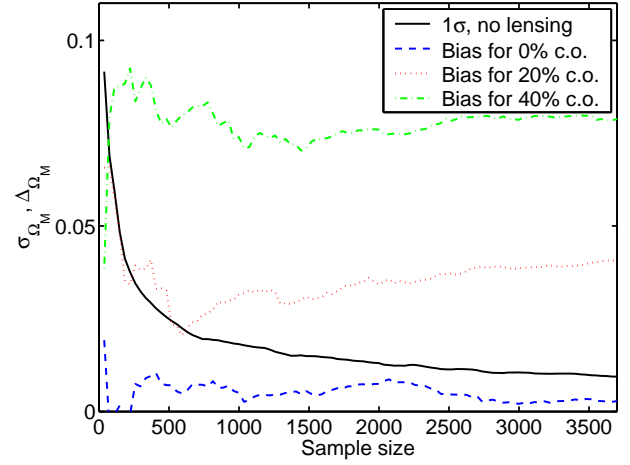,width=.45\textwidth}}}
  \caption{The evolution of the systematic error due to
    gravitational lensing for different dark matter models and the
    statistical error (the solid curve) as a function of the sample
    size. Note that zero SNe correspond to zero high-$z$ events, i.e.,
    the fits are only based on the 300~low-$z$ events that are always
    added. The scatter in the lensing bias curves illustrates the
    stochastic nature of the effect: a handful of high-magnification
    supernovae may produce large effects in the fitted cosmological
    parameters.}
  \label{fig:statsys}
\end{figure}

In Fig.~\ref{fig:statsys} the solid curve is the estimated statistical
error of $\om$ plotted as a function of the sample size. It should be
noted here that zero events means zero events from the high-$z$
distribution, whereas the 300~events from the Supernova Factory
simulations are always present.
It is interesting to note in this figure that the full data set of
2300 SNe is not necessary to get systematic errors comparable to the
statistical error, but gravitational lensing has to be considered also
for rather small data samples containing high redshift SNe ($z\gsim1$)
if the fraction of compact objects is non-negligible.

\section{Determining the lensing dispersion}\label{sec:disp}
As we have seen, gravitational lensing may induce sizeable systematic
effects when trying to
determine cosmological parameters using the distance-redshift relation
for standard candle sources.  A virtue of gravitational lensing is
however that the distribution of luminosities can be used to obtain
information on the matter distribution in the Universe, e.g., to
determine the fraction of compact objects like primordial black holes
or MACHOs.

As shown in \citet{fraction}, gravitational lensing effects are quite
similar for different smooth dark matter halo distributions, but very
sensitive to the fraction of the matter density in compact objects,
$f_{\rm p}$. Assuming that the major part of the matter density in the
universe is in either smooth halos or compact objects, we can thus
parametrize lensing effects with $f_{\rm p}$.

\begin{figure}[thb]
  \centerline{\hbox{\epsfig{figure=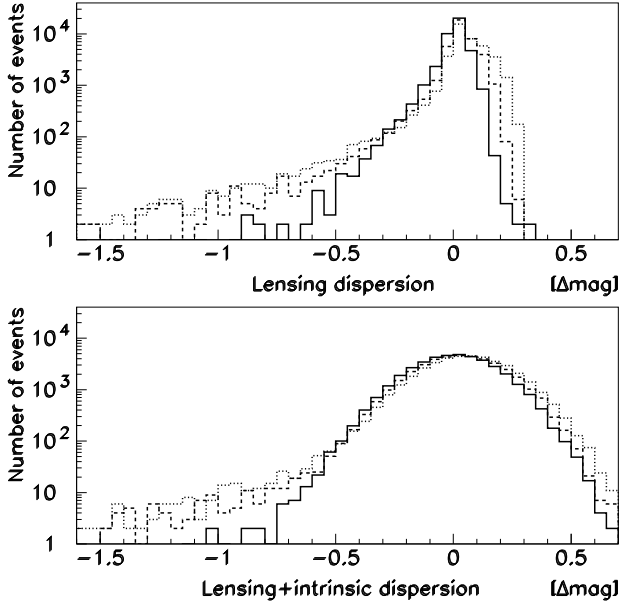,width=.5\textwidth}}}
  \caption{Magnitude dispersion of data set~B for 0 \% (full line), 
    20 \% (dashed line) and 40 \% (dotted line) compact objects using
    logarithmic scale.  The bottom panel includes a Gaussian smearing,
    $\sigma_{m}=0.16$ mag, due to intrinsic brightness differences
    between SNe and from measurement error.}
  \label{fig:biasart1x2}  
\end{figure}

Using SNOC, large data sets of synthetic SN observations over a broad
redshift range were created with a variable fraction of compact
objects ranging from 0 to 40\ts\%, see Fig.~\ref{fig:biasart1x2}.
Using the Kolmogorov-Smirnov test, it was shown how the matter density
in compact objects can be determined using SN data comparable to
set~A or B to $<5$\ts\% accuracy which we will assume to be the
predicted uncertainty in the parameter $f_{\rm p}$, see also 
\citet{metcalf} and \citet{seljak}. 
We now want to parametrize the magnitude distributions obtained from
the Monte-Carlo simulations to obtain approximately correct pdf's to
be used in the
\ml-analysis.

\section{Parametrizing the pdf}\label{sec:para}
We parametrize the pdf's for different fractions of compact objects
and redshifts with a Gaussian with a high magnification tail, i.e.,
\begin{eqnarray}
  \label{eq:fit2}
  f(m)&=&a\cdot e^{-\frac{(m-m_0)^2}{2\sigma ^2}}, \quad m>m_{\rm c}\nonumber \\
  f(m)&=&a\cdot e^{-\frac{(m-m_0)^2}{2\sigma ^2}}+b\cdot |m|\cdot 10^{s\cdot m}, \quad m<m_c.
\end{eqnarray}
This distribution is motivated by the fact that the dominating
intrinsic Gaussian dispersion will be shifted toward fainter values
since a majority of lines-of-sight contain less matter than in a
Friedmann-Lema\^{\i}tre universe. The tail represents the effect of,
e.g., galaxies lying close to some lines-of-sight causing high
magnifications.

We can set $s=1.5$ and $m_c=0$ for all cases and still have a
reasonable $\chisq$. Since the fitted probability density functions
are normalized to give an integrated probability of unity before
performing the maximum likelihood analysis, our parameters are $m_0,
\sigma$ and $b/a$.

\begin{figure}[thb]
  \centerline{\hbox{\epsfig{figure=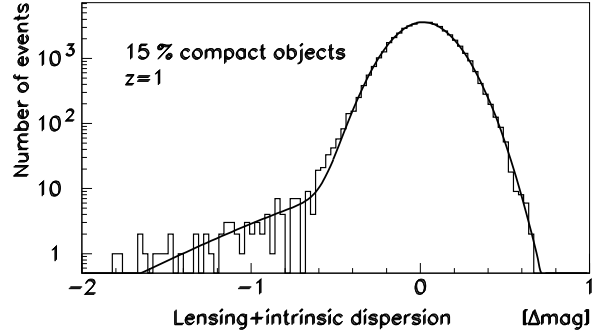,width=0.5\textwidth}}}
  \caption{Fitting Eq.~(\ref{eq:fit2}) to the reference sample
  corresponding to 15 \% compact objects. In this fit, $a=3597,
  m_0=0.01756, \sigma=0.1653$ and $b=90.99$.}
  \label{fig:biasart1x1}
\end{figure}

As an example, see Fig.~\ref{fig:biasart1x1}, where
Eq.~(\ref{eq:fit2}) is fitted to a reference sample corresponding to
40\,000 sources at $z=1$ in a universe with a 15 \% mass fraction in
compact objects. Here, $m_0=0.01756, \sigma=0.1653$ and $b/a=0.02530$
with $\chisq/{\rm ndf}=86.44/68$.

For redshifts $z<0.5$ (regardless of the value of $f_{\rm p}$)  
and $0.5<z<1$ with low $f_{\rm p}$, one may ignore the
high magnification tail (using a simple Gaussian), and still have
acceptable $\chisq$. Note however that the mean is still shifted from
the zero value. For higher $z$ and $f_{\rm p}$, we need the full pdf. E.g.,
for $z=2$ and 20 \% compact objects, we get $\chisq/{\rm ndf}\sim 10$
for a Gaussian and $\chisq/{\rm ndf}\sim 1$ including the high
magnification tail.

\section{Correcting for the bias}\label{sec:corr}
\begin{figure}[thb]
  \centering \subfigure[The fits in the $(\om,\ola)$-plane.]{%
  \hbox{\epsfig{figure=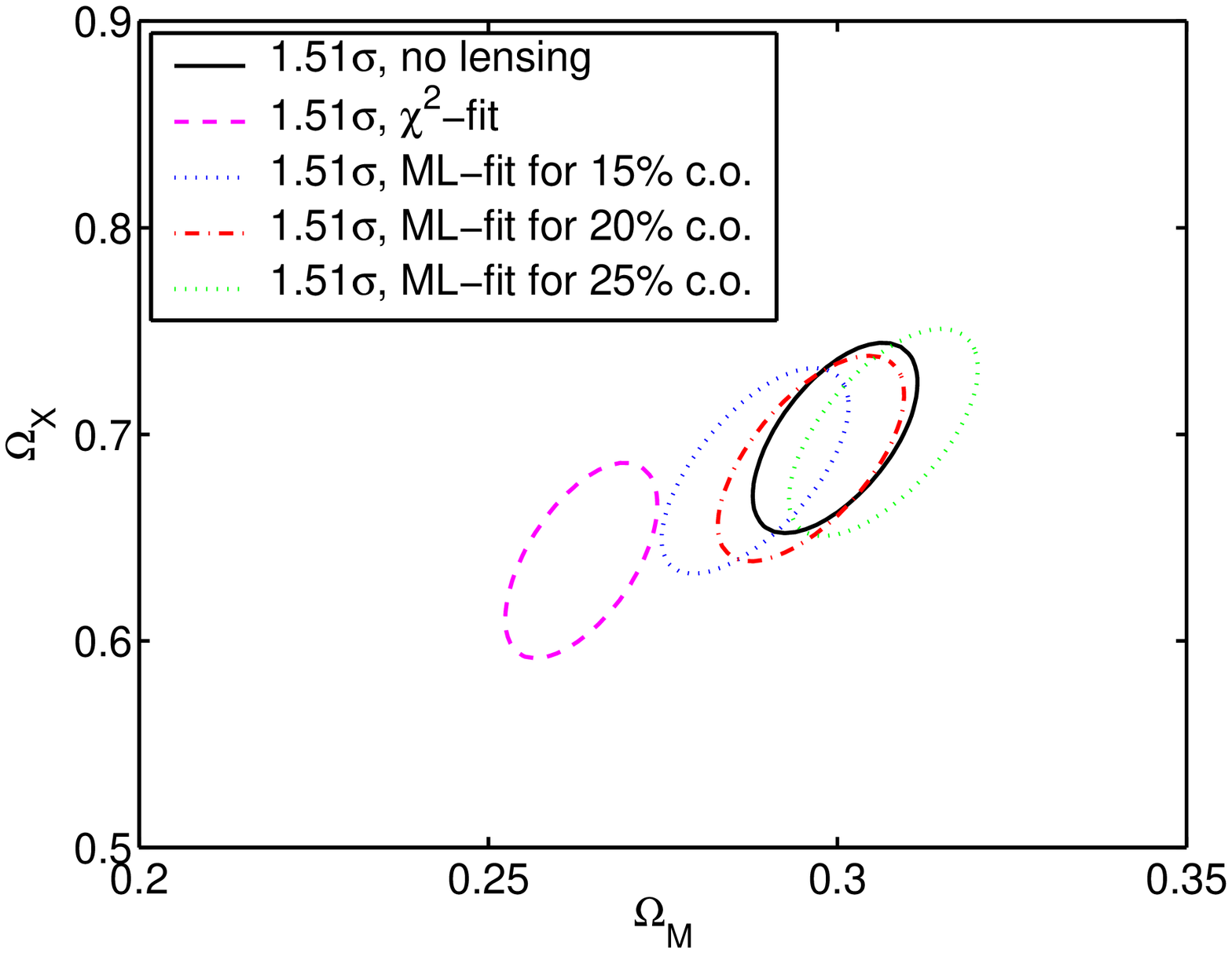,width=.45\textwidth}}}\\
  \subfigure[The fits in the $(\om,w)$-plane, assuming a flat
  universe.]{%
  \hbox{\epsfig{figure=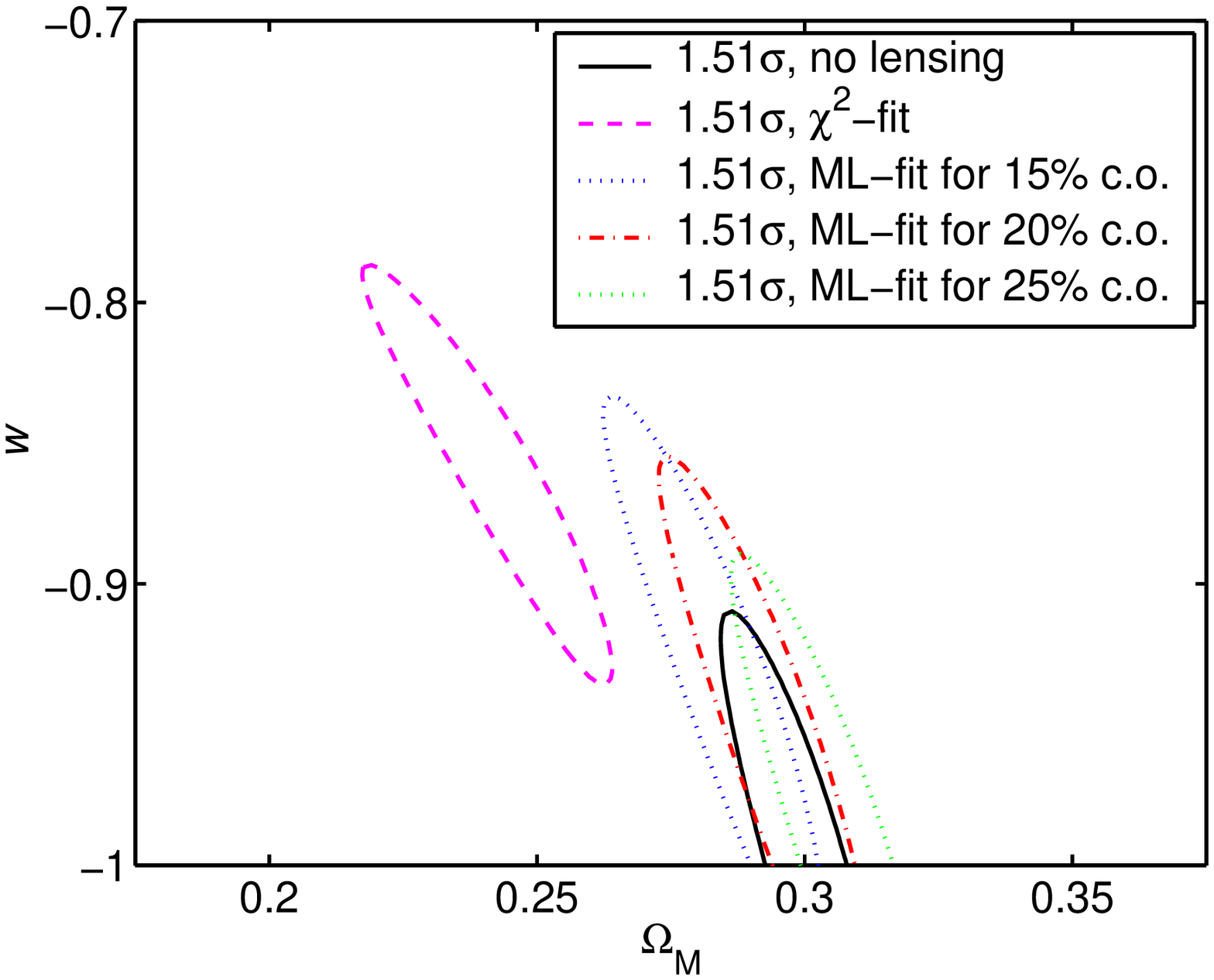,width=.45\textwidth}}}
  \caption{Both plots show the result of three-parameter fits where
  $\sm$ has been fitted as a nuisance parameter. The solid contours
  are the $\chisq$-fit result when lensing effects are absent and the
  dashed contours are the $\chisq$-fit for a dark matter model with
  20\ts\% compact objects and 80\ts\% parametrized as {\nfw} if
  lensing effects are neglected in the fit (cf.
  Fig.~\ref{fig:bias}). The dash-dotted contours show the fit result
  on the same sample when the parametrized pdf presented
  in~Sect.~\ref{sec:para} is used and the correct value of 20\ts\%
  compact objects are used.  The dotted contours show the result when
  an incorrect value ($\pm5\,\%$) of the fraction of compact objects
  is assumed. Note that the confidence regions represent the outcome
  of one specific experimental realisation.}  \label{fig:lenscorr}
\end{figure}

There is apparently a great need to be able to handle gravitational
lensing effects in the fits of the cosmological parameters. If the
magnitude distribution presented in Sect.~\ref{sec:para} is used in a
\ml-analysis the results presented in Fig.~\ref{fig:lenscorr} are
obtained. The contours represent the $\sim68\,\%$ confidence region
for one specific experimental realisation.\footnote{A different
  realisation would give contours of the same size but these may not
  have the same centres. They should however also include the true
  value which gives a limit on the how much of a difference there
  could be between different realisations.} The solid contours are as
previously the
confidence region when no lensing effects have been considered while
the dashed curve is the $\chisq$-fit of for a dark matter model of
20\ts\% compact objects (cf. Fig.~\ref{fig:bias}). The dash-dotted
curves show the \ml-fits using the correct pdf, which is reducing the
bias significantly as compared with the simple $\chisq$-fit. The
dotted curves show the result of an \ml-fit when fractions of 15\ts\%
and 25\ts\% compact objects are assumed when the true value is
20\ts\%. These contours represent the expected systematic bias when
under- or overestimating the fraction of dark matter in compact
objects with 5\ts\%. The results are a big improvement over the
$\chisq$-analysis. Thus, it should be possible to reduce the bias due
to lensing in a fairly effective way, even with the simple approach
used in this note.

\section{Summary and discussion}
Neglecting gravitational lensing effects may cause systematic
errors when determining cosmological parameters using the
distance-redshift relation for Type Ia SNe. E.g., if the universe
contains a large fraction of compact objects, the matter density $\om$
can be severely underestimated or a cosmological constant may be
wrongly excluded if one assumes a Gaussian distribution of
magnitudes. Using an approximately correct pdf for the SNe magnitudes,
this bias can be reduced significantly.

In this note, we show that lensing effects can be parametrized in a
simple way by the fraction of compact objects, $f_{\rm p}$, and the
redshift. In general, lensing effects are bigger for larger values of
$f_{\rm p}$ and higher redshifts. If $f_{\rm p}$ can be determined
with $<5$\ts\% accuracy, the bias in the cosmological parameter
determination can be effectively reduced as compared to the case where
the magnitudes are assumed to be Gaussian distributed.  For future
surveys, e.g., by the proposed {\snap} satellite aiming at doing
precision cosmology with Type Ia SNe, this reduction is essential to
get systematic errors comparable to the statistical errors.

For the simple analysis performed in this note, we have first
determined $f_{\rm p}$ to be able to choose the correct pdf for the
subsequent parameter determination. Another possibility is to make a
simultaneous fit of, e.g., $(f_{\rm p}, \om, \ola, \sm)$ which will be
the aim of some forthcoming work.

\section*{Acknowledgements}
AG is
a Royal Swedish Academy Research Fellow supported by a grant from the
Knut and Alice Wallenberg Foundation. We would like to thank Greg
Aldering for pointing out an assumed error in the redshift range for
Supernova Factory.

\end{document}